*Article (total number of words 6 121, tables 1, figures 7)*

# Silver-enriched Microdomain Patterns as Advanced Bactericidal Coatings for Polymer-based Medical Devices

*Jana Pryjmaková [1], Barbora Vokatá [2], Miroslav Šlouf [3], Tomáš Hubáček [4], Patricia Martínez-García [5,6], Esther Rebollar [7], Petr Slepička[1] and Jakub Siegel [1,\**

[1] Department of Solid-State Engineering, University of Chemistry and Technology Prague, Technická 5, 166 28 Prague, Czech Republic; jana.pryjmakova@vscht.cz (J.P.), petr.slepicka@vscht.cz (P.S.), jakub.siegel@vscht.cz (J.S.)

[2] Department of Microbiology, University of Chemistry and Technology Prague, 166 28 Prague, Czech Republic; vokataa@vscht.cz (B.V.)

[3] Institute of Macromolecular Chemistry, Academy of Sciences of the Czech Republic, Heyrovského nám. 2, 162 06 Prague, Czech Republic; slouf@imc.cas.cz (M.Š.)

[4] Biology Centre of the Czech Academy of Sciences, SoWa National Research Infrastructure, Na Sádkách 7, 370 05 České Budějovice, Czech Republic; hubacektom@gmail.com (T.H.)

[5] Depto. Física Interdisciplinar, Universidad Nacional de Educación a Distancia (UNED), 28232 Las Rozas de Madrid, Spain; patricia.martinez@ccia.uned.es (P.M.-G.)

[6] NANOesMAT, UNED, Unidad Asociada al CSIC por el IEM y el IQF, 28232 Las Rozas de Madrid, Spain

[7] Instituto de Química Física Blas Cabrera, IQF-CSIC, Calle de Serrano 119, 28006 Madrid, Spain, e.rebollar@csic.es (E.R.)

\* Correspondence: jakub.siegel@vscht.cz; Tel.: +420 220 445 149

**Abstract:** Today, it would be difficult for us to live a full life without polymers, especially in medicine, where its applicability is constantly expanding, giving satisfactory results without any harm effects on health. This study focused on the formation of hexagonal domains doped with AgNPs using a KrF excimer laser (λ=248 nm) on the polyetheretherketone (PEEK) surface that acts as an unfailing source of the antibacterial agent - silver. The hexagonal structure was formed with a grid placed in front of the incident laser beam. Surfaces with immobilized silver nanoparticles (AgNPs) were observed by AFM and SEM. Changes in surface chemistry were studied by XPS. To determine the concentration of released $Ag^+$ ions, ICP-MS analysis was used. The antibacterial tests proved the antibacterial efficacy of Ag-doped PEEK composites against *Escherichia coli* and *Staphylococcus aureus* as the most common pathogens. Because AgNPs are also known for their strong toxicity, we also included cytotoxicity tests in this study. The findings presented here contribute to the advancement of materials design in the biomedical field, offering a novel starting point for combating bacterial infections through the innovative integration of AgNPs into inert synthetic polymers.

## 1. Introduction

Synthetically prepared polymers are an integral part of the medical and pharmaceutical industries because of their unmistakable physical and chemical properties. Their range of applications is very wide; from medical packaging through cover and encapsulating devices to implants, prosthetics, and drug carriers themselves. In addition to physical and chemical properties, polymers are valued for their ease of shaping, allowing for the creation of compositions of any shape and size.

When the polymer medical device comes into contact with the tissue, numerous processes may occur that are not always desirable. Infections [1], inflammations [2], immune reactions [3] or osteolysis [4] can cause implant failure and health problems. In particular, polymeric medical devices are generally prone to degradation, fragility, and biofilm formation. These undesirable side effects are due to the influence of the biological environment. To overcome this, the polymeric material can be improved by adding inorganic nanoparticles that provide tensile strength [5], mechanical strength [6], and antibacterial effect with current wound healing capabilities [6].

Polyetheretherketone (PEEK) is a high-performance polymer that has gained significant attention and application in various medical fields, mostly in the form of composites. In dentistry, PEEK has found its place as a biomaterial for dental prosthetics, including crowns, bridges, and removable dentures due to its biocompatibility, excellent mechanical strength, and resistance to wear and corrosion [7-9]. The ability to mimic the mechanical properties of bone enhances osseointegration [10], making PEEK popular for the manufacture of implants such as joint replacements, plates, and screws [11-13]. PEEK implants also exhibit a lower modulus of elasticity compared to traditional metallic implants (Ti6Al4V), which is closer to the modulus of cortical bone [14]. Furthermore, PEEK-based implants do not change their mechanical properties after sterilization with gamma rays or ethylene oxide [15, 16].

In the dynamic landscape of biomedical materials, the integration of silver nanoparticles (AgNPs) holds immense promise in imparting antimicrobial properties to polymers, thus improving their suitability for various biomedical applications [17-19]. Recent advances in nanotechnology and material science have paved the way for novel strategies in the synthesis of polymer composites with precisely immobilized AgNPs, opening avenues for improved biocompatibility and antibacterial efficacy.

Currently, the number of studies dealing with Ag-PEEK-based composites is constantly increasing, which expands the knowledge in the field of materials engineering and their potential applications in medicine. Ag-PEEK composites represent excellent materials with antibacterial efficacy [20]. The challenging task is to prepare the Ag-PEEK composite with antibacterial properties that do not exclude the biocompatibility. The study of Jiang et al. [21] presented an Ag-PEEK-based hydrogel with antibacterial efficacy against *E. coli* and *S. aureus* using an ultrasonic and mixing technique. Furthermore, the hydrogel showed enhanced pre-osteoblast cell viability. In addition, 3D printed PEEK implants have been applied in biomedical applications. Deng et al. [22] developed a 3D printed PEEK implant for bone repair with functionalized AgNPs via chemical reduction. This implant also exhibited antibacterial activity and provided cell proliferation.

The preparation of AgNP-polymer composites and the attachment of the particles itself is a hot topic. There are several ways to prepare such composites, among which we can include chemical grafting of nanoparticles [23, 24], mixing them into a polymer solution [25, 26] or laser-assisted immobilization. The advantage of laser immobilization lies in the fast and firm anchoring of nanoparticles without the need for chemical bonds. Electrochemically prepared silver nanoparticles can be immobilized, eliminating complex chemical synthesis, which mostly requires a significant amount of chemicals and complicated preparation conditions (heating, cooling, long synthesis time, etc.). This study focuses on the laser immobilization of AgNPs within polyetheretherketone (PEEK) matrices, encapsulating nanoparticles into hexagonal domains with enhanced stability and controlled distribution. The judicious combination of polymers and AgNPs offers a synergistic approach that leverages the inherent properties of both components for tailored biomedical applications. The laser immobilization technique serves as a key methodology for achieving precise control over the incorporation of AgNPs within the polymer matrix [27]. This process not only ensures a uniform distribution of AgNPs in very surface, but also allows the formation of different-shaped structures [28], offering a unique structural configuration that has the potential to optimize the antibacterial properties of the resulting material.

The significance of this research lies in its potential to advance the field of biomaterials by providing a method for engineering polymer composites with enhanced antibacterial efficacy and biocompatibility. The results of this study may find applications in various biomedical settings, ranging from implantable medical devices to tissue engineering, where the need for materials with superior antimicrobial properties is paramount. Throughout this article, we will represent the laser immobilization process leading to the design of the PEEK surface with Ag-enriched domains, present the characterization of the synthesized materials, and discuss the implications of our findings in the broader context of biomedical research and applications.

**2. Materials and Methods**

*2.1 General Materials*

For the electrochemical synthesis of Ag nanoparticles (AgNPs), two silver electrode ($40\times10\times1$ mm$^3$, purity 99.99%, Safina a.s., Vestec, Czech Republic) and a solution of sodium citrate dihydrate ($Na_3C_6H_5O_7$, Sigma–Aldrich Co., St. Louis, MO, USA) were used. The AgNPs were immobilized into surface of polyetheretherketone foil (PEEK) obtained from Goodfellow Cambridge Ltd., Huntingdon, UK with density of 1.3 g·cm$^{-3}$, and thickness of 50 µm. Antibacterial efficacy was studied against *Escherichia coli* (*E. coli*; CCM 4517, Czech Collection of Microorganisms, Bohunice, Czech Republic) and *Staphylococcus aureus* (*S. aureus*; CCM 4516) and toxicity on primary lung fibroblasts (MRC-5, American Tissue Culture Collection, Manassas, VA, USA). Minimal Essential Medium Eagel (MEM) and L-Glutamine were obtained from Sigma-Aldrich, St. Louis, MO, USA and fetal bovine serum (FBS) was obtained from Thermo Fisher Scientific, Waltham, MA, USA.

*2.2 Synthesis of Ag Nanoparticles and Sample Preparation*

In this work, the colloid AgNPs were prepared by electrochemical synthesis using silver electrodes. The silver electrodes were immersed in a sodium citrate solution (volume 120 ml, concentration 1 mmol·l$^{-1}$). The electrodes were powered by 15 V and the solution was stirred vigorously for 20 min. The solution was then decanted, filtered and kept in a dark place for 24 h. The next day, the concentration of AgNPs was determined using AAS analysis. The size of the synthesized nanoparticles was approximately 25 nm with stability of 3 days. The AgNPs prepared were then immobilized into the surface of the polyetheretherketone foil (PEEK) using a KrF excimer laser (COMPex PRO 50 F, Coherent Inc., Santa Clara, CA, USA). The PEEK foil was placed in a cuvette filled with colloid AgNPs and irradiated with linearly polarized light (248 nm) with laser fluence of 10 mJ·cm$^{-2}$. Irradiation was accomplished at an incidence angle of 90° with 6000 pulses. The method of preparation of AgNPs and their subsequent immobilization under set conditions are described in detail in our previous study [29]. In this way, the samples with a planar surface seeded with AgNPs (Ag/PEEK) were prepared. To achieve hexagonal domains on the PEEK surface, an aluminum grid ($r_{min}$= 200 µm, $r_{max}$= 231 µm, Liss a.s., Rožnov pod Radhoštěm, Czech Republic) was placed behind the aperture (see Scheme 1). Samples with hexagonal domains are denoted as #Ag/PEEK.

*2.3 Analytical Methods*

The concentration of colloid AgNPs was determined using atomic spectrometry (AAS) on the Agilent 280FS AA flame atomizer spectrometer (Agilent Technologies, Tokyo, Japan).

Initially, the hexagonal structure formed on the PEEK foil surface was observed using the Olympus LEXT OLS3100 confocal laser scanning microscope (Olympus Corporation, Tokyo, Japan) equipped with a GaN laser (wavelength 405 nm, dark field mode, objective 5x).

The information about the surface morphology and its parameters was studied with atomic force microscopy (AFM). The samples were scanned using a Dimension ICON device (Bruker Corp., Billeria, MA, USA) in ScanAsyst® mode. The AFM was equipped with a SCANASYST-AIR probe with a silicon tip attached to the SiN cantilever (elasticity of 0.4 N·m$^{-1}$, natural frequency of 70 kHz, scan rate 0.5 Hz, Bruker corp., USA). Data about surface morphology and parameters such as surface roughness ($R_a$) and the value of the surface area difference (SAD) [29] were evaluated using the NanoScope® v1.8 analysis program (Bruker Corp., Billerica, MA, USA).

Scanning electron microscopy (SEM) was used for a more detailed view of the resulting surfaces of samples embedded with nanoparticles. The visualization was performed on a MAIA 3 field emission gun scanning microscope (FEG-SEM, high-resolution mode, secondary electron detector, accelerating voltage of 3 kV, TESCAN, Brno, Czech Republic). Samples were fixed in place with double-sided carbon tape (Cristine Groepl, Tulln, Austria) and coated with a layer of carbon in a JEE-4C vaporizer (JEOL, Akishima, Japan).

Surface chemistry was analyzed with an ASCEProbeP Omicron nanotechnology X-ray photoelectron spectroscope (Scienta Omicron, Taunusstein, Germany). The source was monochromatic X-rays with an energy of 1486.7 eV, the chamber pressure was $2 \times 10^{-8}$ Pa. Electrons were collected at take-off angles of 90 ° (perpendicularly to the sample surface) and 9 °.

To determine the wettability of the composite surface, the contact angle (CA) of the water droplet was measured using the sessile drop method on a KRÜSS DSA 100 goniometer (KRÜSS, Hamburg, Germany) equipped with an automatic pipette. The evaluation of the data obtained was performed using a KRÜSS Advance software v2.0. CAs were calculated using the three-point method for ten droplets of distilled water (volume of 2 μl). Then the arithmetical means and standard deviations were calculated, which are expressed in form (A±B).

When inductively coupled plasma mass spectroscopy (ICP-MS) of the leachates was performed, it was determined whether and in what quantity silver ions are released from the samples into the environment, which information was important for the evaluation of the antibacterial effect. To determine the concentrations of the released ions, leachates were prepared under the same conditions as in the antibacterial tests. The samples were immersed in 5 ml of distilled water for 3 and 24 h. As control samples, distilled water without polymer samples was used. Before analysis, each leach was centrifuged for 30 minutes with the Optima MAX-XP Ultracentrifuge (Beckman Coulter, Brea, CA, USA) at a maximum load of 200,000 g to eliminate the final residue of solid metals. Ag$^+$ ions concentrations were measured using an Agilent 8800 triple-quadrupole spectrometer (Agilent Technologies, Santa Clara, CA, USA) with an autosampler. The samples were injected into the atomizer using a MicroMist device with a peristaltic pump. Then the arithmetical means and standard deviations were calculated, which are expressed in form (A±B).

*2.4 Biological Tests*

The investigation of the antibacterial properties of #Ag/PEEK was performed using drop plate tests. PEEK was used as the control sample. The tests are based on the counting of viable bacteria [30]. In this work, the bactericidal effect was tested against Gram-negative (G$^-$) *Escherichia coli* (*E. coli*; CCM 4517) and Gram-positive (G$^+$) *Staphylococcus aureus* (*S. aureus*; CCM 4516). Using an experimental setup [29], the colony-forming units (CFUs) of the bacteria were counted, the arithmetical means and standard deviations were calculated.

The cytotoxic effect of #Ag/PEEK was observed on primary lung fibroblasts (MRC-5). They were cultured in Minimal Essential Medium (MEM) supplemented with 2 mM L-Glutamine and 10% fetal bovine serum under specific conditions (37 °C, 5% CO2 and 95% humidity). For cell viability measurement, a resazurin assay was used [31]. The samples were inserted into the culture plate (6-well, VWR, Radnor, PA, USA), sterilized with 70% ethanol (purity of 99.9%) for 30 min and washed with phosphate buffer (PBS, pH 7.4). Subsequently, the samples were inoculated with cultured cells (30,000 cells/cm$^{-2}$ in 2.5 mL) and cultivated for 24 and 48 h. Cells samples were then removed from the medium, washed with PBS and incubated with a resazurin solution (final concentration 25 g/mL) in a MEM medium without phenol red for 4 h. The fluorescence measurement was performed on Fluoroskan Ascent (Thermo Labsystems, Waltham, MA, USA) and excitation and emission wavelengths of 560 and 590 nm, respectively. Cell viability was represented as a percentage of metabolic activity of control cells grown on standard tissue culture polystyrene (TCPS). Mean values and standard deviations were calculated.

$$viabillity\ of\ cells\ (\%) = \frac{OD\ treatment}{OD\ control} \times 100 \qquad (1)$$

All samples for drop plate tests and cytotoxicity tests were prepared in triplicate. All experiments were carried out under sterile conditions. The data obtained were processed as graphs with error bars.

**3. Results and Discussion**

*3.1. Nanoparticle Synthesis*

For the repetitive procedure, the concentrations of colloid AgNPs were determined using AAS. Typically, the concentration values were in the range of 25 – 40 mg·l$^{-1}$. Such a difference can be caused by many factors such as uneven immersion of electrodes, nonparallel electrode placement, uneven temperature of the electrolyte, or nuances in the concentration of sodium citrate solution. For this reason, colloid solutions were prepared in triplicates, and those with a concentration exceeding 30 mg·l$^{-1}$ were diluted to 30 mg·l$^{-1}$.

*3.2 Surface Characterization*

An initial overview of AgNPs-enriched domains was performed using a confocal laser scanning microscope (CLSM). The PEEK decorated with hexagonal structures is shown in Figure 1. One can see that the immobilization was successful, and that hexagonal domains were presented. Interestingly, in the middle of domains line patterns appeared, the spacing of which was increased towards the edge of the domain.

The surface morphology of AgNPs immobilized PEEK was also studied using atomic force microscopy (AFM). Scans of Ag/PEEK and #Ag/PEEK are shown in Figure 2. The corresponding surface parameters, such as average surface roughness $R_a$ and surface area difference SAD, are summarized in Table 1. Irradiation of polymer substrates with polarized light from excimer lasers leads to the formation of nanostructures that depend on the modification process parameters. Laser irradiation in the air can form globular [32] or ripple-like [33, 34] nanostructures. However, laser modification of polymers is possible even in water or liquid solutions [35]. When appropriate mesh is used, the resulting nanostructures are enclosed within the hexagonal domains. In an overview 3 µm scan (Figure 2 – #Ag/PEEK), the surface of PEEK with immobilized AgNPs appears planar (Figure 2 – Ag/PEEK), however, upon closer inspection, we found that there are globular formations on the surface. This is also evidenced by the increase in the SAD value. As PEEK was modified in solution of colloid AgNPs, the nanoparticles presented in each globule are clearly visible in Figure 2b. Since we used an immobilization laser fluence of 10 mJ·cm$^{-2}$, which is optimized for planar surface with immobilized AgNPs [29], the presence of globular structures can be explained as the result of the combination of the effects of constructive interference of the laser beam on the grid wounds with reflection of radiation from the polished surfaces of the grid during the passage of radiation, which can locally increase the applied fluence over the integral value (Figure 1). Furthermore, the surface morphology changed within different observed areas, which is clearly seen in the SEM images.

The closer investigation of the #Ag/PEEK surface with scanning electron microscopy (FEG-SEM) brings detailed information about the formed nanostructures. Micrographs of the different areas on the surface of #Ag/PEEK can be seen in Figure 3. The surface was observed in 3 distinctive areas on the sample: between (b), on the edge (c), and in the middle of the domains (d). In Figure 3a, domains with hexagonal shape doped with Ag nanoparticles are clearly seen, representing a light gray color. The most striking observation that emerged from the SEM analysis was a different surface morphology in the middle of the domain (Figure 3b). Our expectation was the presence of the planar surface with immobilized nanoparticles, although the presence of typical lines within the hexagonal domains during the initial confocal microscopy screening suggested that the structure may be more complicated. However, the LIPSS formation occurred in the form of ripples, which is quite interesting. LIPSS on the polymer surface can be generated using higher laser fluences, typically ranging from 18 – 30 mJ·cm$^{-2}$ [36], but in the case of the present study, the LIPSS began to form at 10 mJ·cm$^{-2}$. The answer to this phenomenon can be found in the interaction of laser beam when passing through the grid, similarly as in the case of globular nanostructure formation. As the samples were irradiated with a laser beam through the grid, constructive interference of the laser beam on the grid wounds with reflection of radiation from the polished surfaces of the grid during the passage of radiation may occur, which may locally increase applied fluence over the integral value [37, 38]. If we were to accept this interpretation, the most significant morphological deviations from the planar structure would be observable toward the center of the hexagonal structures, which was confirmed by both AFM and SEM data. On the contrary, at the edge of the domains (Figure 3c), the planar surface seeded with AgNPs was obvious. Finally, the area between domains contained an order of magnitude fewer nanoparticles, as a result of the optical phenomena mentioned, when the grid used cannot fully shield the penetrating radiation (Figure 3d). However, this area can serve as a space for the possible attachment of cells.

In addition to the change in surface morphology, there was a change in the surface chemistry of the prepared composites. Data obtained from XPS are summarized in Figure 4. Compared to the C and O content in PEEK, there was a decrease in both elements for both take-off angles, as expected. The presence of Ag confirms the successful implementation of AgNPs. At a take-off angle of 9°, the Ag concentration reached approximately 18 at.%, which is two times higher compared to the 90 °. This dramatic increase is related to the angle-resolved XPS (AR-XPS) analysis, when the analytical information at 90 ° is of 8 to 10 nm depth, while the information obtained at 9 ° is much more superficial (1-2 nm) [39]. This increase is worth highlighting because the vast majority of nanoparticles are immobilized pretty close to the surface and thus are highly bioavailable.

Knowledge of surface wettability is needed for further biological tests as an important parameter for interaction with cells. The measured contact angles (CAs) of PEEK and #Ag/PEEK with the corresponding images of sessile drops can be seen in Figure 4. The hydrophilic character of the PEEK surface is evident in Figure 5. The CA of #Ag/PEEK decreased after surface design with hexagonal domains; however, the hydrophilic character of surface was preserved. These results are not in line with those of Liu et al. [40], who studied the wettability of the PEEK-based composite with sputtered silver nanolayers. After silver sputtering, CA increased significantly from approximately 65 ° to 130°, as the water droplet was in contact with the silver nanolayer only. From this finding it follows that laser immobilization seems to be an effective method of AgNPs immobilization without changes of surface wettability character.

For a subsequent evaluation of the bactericidal properties, the concentration of $Ag^+$ ions released was determined. The concentration values for 3 and 24 h of leaching, which correspond to the incubation times in the case of antibacterial tests, are summarized in Table 2. The highest concentrations $14.5 \times 10^{-3}$ µg·ml$^{-1}$ of $Ag^+$ ions were achieved after 3 and 24 h in the case of Ag/PEEK, since the entire surface (5 × 10 mm$^2$) was seeded with Ag nanoparticles. However, the concentration did not reach the minimal inhibitory concentration (MIC) of *E. coli* after 24 h. Based on a study by Ning et al. [41], the minimum inhibitory concentration for *E. coli* and *S. aureus* is $24 \times 10^{-3}$ µg·ml$^{-1}$. In the case of #Ag/PEEK the concentrations of $Ag^+$ ions were even lower, as between individual domains the AgNPs were present only in a minimal amount. However, it was found that even a small concentration of silver ions may be sufficient to kill bacteria.

*3.2 Biological Tests*

Because PEEK was doped with AgNPs, the antibacterial efficacy of prepared samples was studied using drop plate tests. *E. coli* and *S. aureus* represent one of the most common pathogens in medical device-related infections and antibiotic resistance (*S. aureus*) [42-44]. From Figure 6a, one can see that PEEK did not show any antibacterial effect on *E. coli* at incubation times of 3 and 24 h. After AgNPs immobilization, the slight decrease of CFU of *E. coli* was observed after 3 h of incubation time. Strong evidence of antibacterial efficacy was found after 24 h of incubation time, when *E. coli* growth was completely inhibited. The effectiveness of electrochemically synthesized AgNPs incorporated in the PEEK foil is comparable to that of those prepared via green synthesis. In the study of Lethongkam et al. [45], AgNPs synthesized using an extract of *Eucalyptus camaldulensis* were incorporated into the silicon urinal catheter, expressing an eminent antibacterial effect against *E. coli*. In the case of *S. aureus*, the results were interestingly different. At an incubation time of 3 h, the interaction of *S. aureus* with PEEK foil did not cause inhibition of bacterial growth. The single most striking difference revealed was the dramatic decrease in *S. aureus* CFU after 24 h for PEEK. PEEK samples doped with Ag showed antibacterial efficacy after 24 h of incubation time. As samples for antibacterial tests were prepared as leaches of polymeric samples in medium with bacteria, $Ag^+$ ions play a key role in the antibacterial properties. The $Ag^+$ ions penetrated and destroyed the cell wall of bacteria, bound to proteins and bases in DNA, leading to DNA damage and cell death. Furthermore, silver ions interact with respiratory enzymes that block cell replication [46, 47]. Some similarities can be found in the study of Deng et al. [22], who cultivated bacteria after contact with PEEK samples doped with Ag nanoparticles. In their case, Ag/3D-PEEK also showed an antibacterial effect after 24 h of incubation time.

The material used in bioapplications must not be harmful to cells and must not show side effects on the surrounding environment. For this reason, cytotoxicity tests were performed on primary lung fibroblasts. The results of these tests are presented in Figure 6b. The tests revealed moderate cytotoxicity of #Ag/PEEK compared to the control samples (TCPS). Relative cell viability did not reach 70%, which is the limit of cytotoxicity according to ISO 10993-5. We assumed that the presence of the clear polymer between the hexagonal domains offered a place for cell adhesion without the toxic effect of AgNPs. Furthermore, LIPSS build up in the middle of domains may positively affect cell adhesion and proliferation, as surface roughness and morphology are important parameters in the scaffold-tissue interaction [34, 48, 49]. However, the LIPSSs were present just in the middle of the domains and were fully seeded with AgNPs, so cells came into contact directly with AgNPs, which had a toxic effect on them.

In summary, #Ag/PEEK provided antibacterial efficacy on Gram positive and Gram negative bacteria; however, the cytotoxic effect is presented. For this reason, our future study should focus on the surface morphology, concentration, and depth of the attached nanoparticles. The solution to this problem can be found in the laser treatment of PEEK as in our previous study [29], where PEEK doped with AgNPs was later irradiated with the KrF laser. The LIPSS nanostructures that serve for cell attachment occurred, and AgNPs were incorporated as antibacterial agents hidden under these structures. Other ways can also be pretreatment of PEEK and immobilization of nanoparticles in LIPSS.

## 4. Conclusions

In conclusion, the surface modification of PEEK with hexagonal domains decorated with Ag nanoparticles presents an intriguing avenue for enhancing its properties for medical device applications. Through the use of advanced imaging techniques, such as CLSM, AFM, and FEG-SEM, we have gained valuable insights into the morphology and composition of the modified surfaces. The formation of hexagonal domains and the presence of Ag nanoparticles within these domains have been successfully observed, showcasing the efficacy of the immobilization process. Moreover, the laser irradiation process has induced the formation of surface nanostructures, further enhancing its characteristics for cell attachment. The observed variations in surface morphology and chemistry, particularly the formation of LIPSS and changes in chemical composition, demonstrated the complexity of the modification process and its impact on surface properties.

Importantly, surface modification has not significantly altered the wettability characteristics of PEEK, which is crucial for its interaction with biological entities. Furthermore, the antibacterial efficacy of the modified surfaces has been confirmed, with controlled release of $Ag^+$ ions contributing to their antimicrobial properties. However, cytotoxicity tests have revealed moderate cytotoxic effects of modified surfaces on primary lung fibroblasts, suggesting the need for further optimization to minimize possible adverse effects while maintaining the desired properties. Nonetheless, #Ag/PEEK composite represents promising material for disposable medical devices (urinary, vascular catheters), tissue carriers, medical plastics that suffer from bacterial colonization, or other devices that are integrated into the bodies of patients for a long time. The presence of AgNPs reduces the risk of secondary infections (e.g. CAUTI) and thus significantly shortens the recovery time.

In general, this study underscores the potential of surface modification techniques for tailoring the properties of medical device materials, opening up new possibilities for improving their performance and biocompatibility in clinical settings. Further research in this direction holds promise for advancing the development of safer and more effective medical devices.

**Author Contributions:** Conceptualization, visualization and writing – original draft, J.P.; sample preparation, antibacterial tests, cytotoxicity tests, B.V.; FEG-SEM analysis, M.Š.; ICP-MS analysis T.H.; sample preparation, P.M-G.; students supervision, E.R.; funding acquisition, P.S.; methodology, writing-review and editing, AFM analysis, J.S. All authors have read and agreed to the published version of the manuscript.

**Funding:** This research was funded by Project OP JAK_Mebiosys, No CZ.02.01.01/00/22_008/0004634 of the Ministry of Education, Youth and Sports, which is co-funded by the European Union.

**Data availability statement:**

The data presented in this study are available at https://doi.org/10.5281/zenodo.12582672.

**Figures**

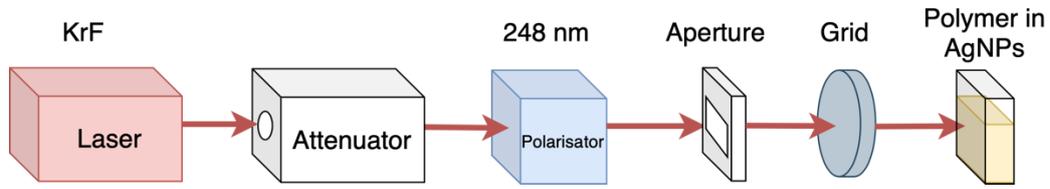

**Scheme 1.** Setup for AgNPs immobilization into polymer foil.

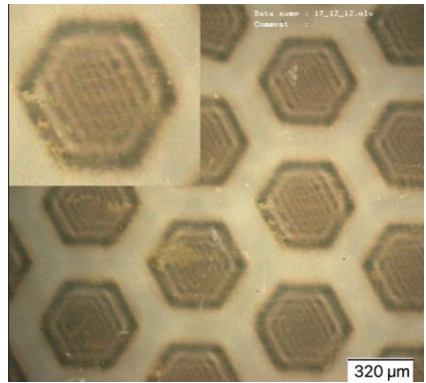

**Figure 1.** Image of hexagonal domains doped with silver nanoparticles obtained from confocal laser scanning microscopy.

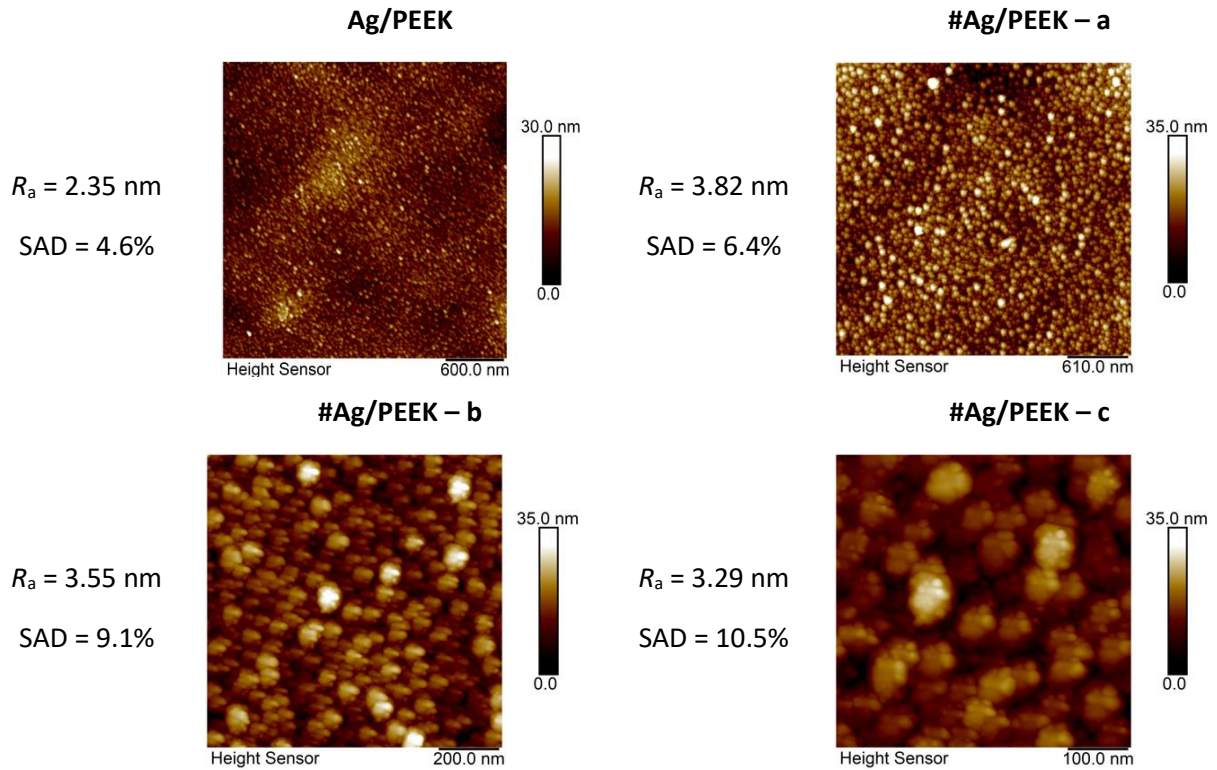

**Figure 2.** AFM scans of Ag/PEEK and #Ag/PEEK with scanning area (a) 3 µm, (b) 1 µm, (c) 500 nm.

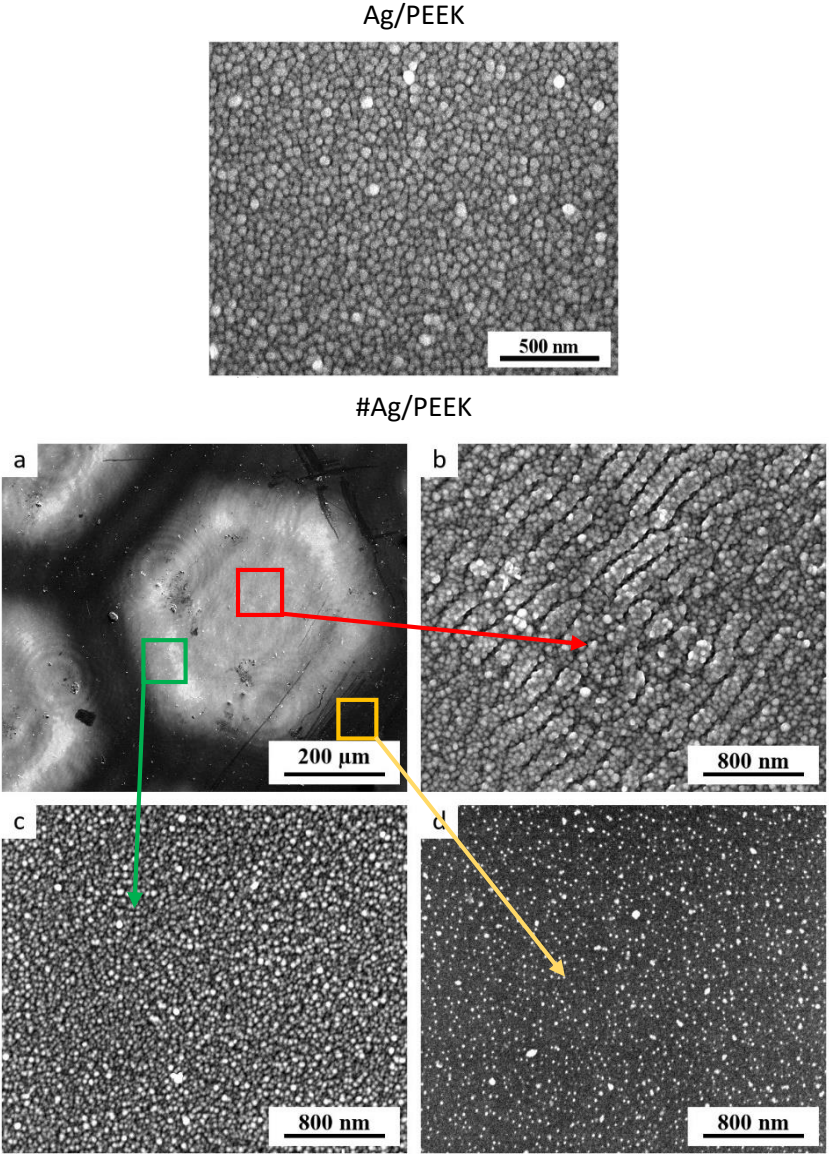

**Figure 3.** Micrographs of Ag/PEEK and #Ag/PEEK with a focus on (a) hexagonal domain, (b) the LIPSS in the middle of the domain, (c) AgNPs in the edge, and (d) between domains obtained from FEG-SEM.

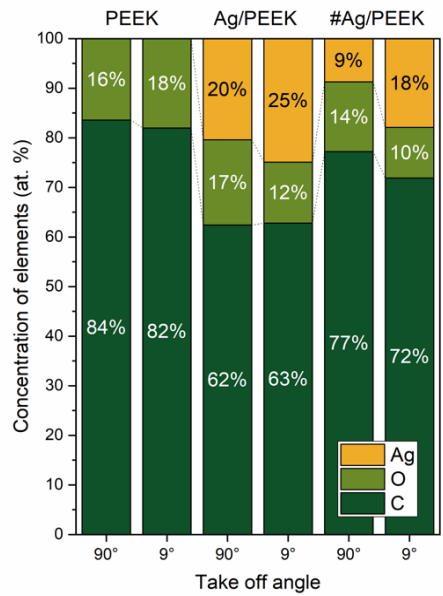

**Figure 4.** Concentration of elements carbon (C), oxygen (O), and silver (Ag) present in the surface of PEEK, Ag/PEEK and #Ag/PEEK for take-off angles 90 ° and 9 °.

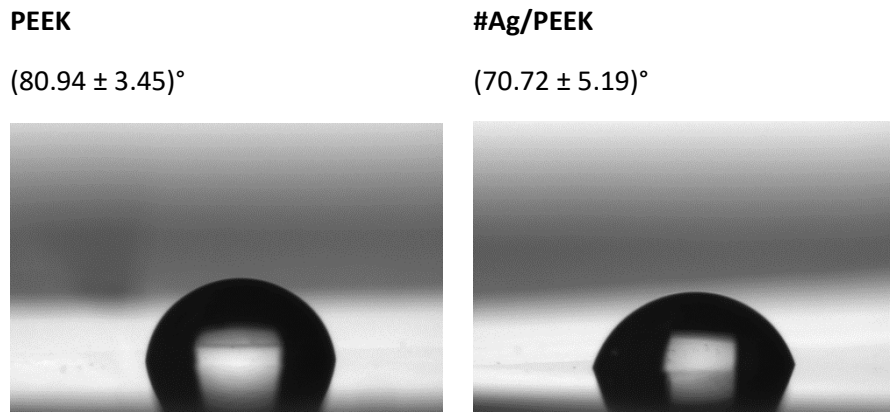

**Figure 5.** Images of water droplets on the surface of PEEK and #Ag/PEEK with the corresponding contact angles. Values represent means ± standard deviations.

**Table 1.** Concentration of $Ag^+$ ions released from PEEK, Ag/PEEK, and #Ag/PEEK after incubation times 3 and 24 h. Values represent means ± standard deviations.

| ($\mu g \cdot l^{-1}$) | 3 h | 24 h |
|---|---|---|
| Control | 0 ± 0.2 | 0 ± 0.1 |
| PEEK | 0.1 ± 0.1 | 0.1 ± 0.2 |
| Ag/PEEK | 14.5 ± 0.2 | 18.7 ± 0.1 |
| #Ag/PEEK | 2.5 ± 0.5 | 7.8 ± 1.6 |

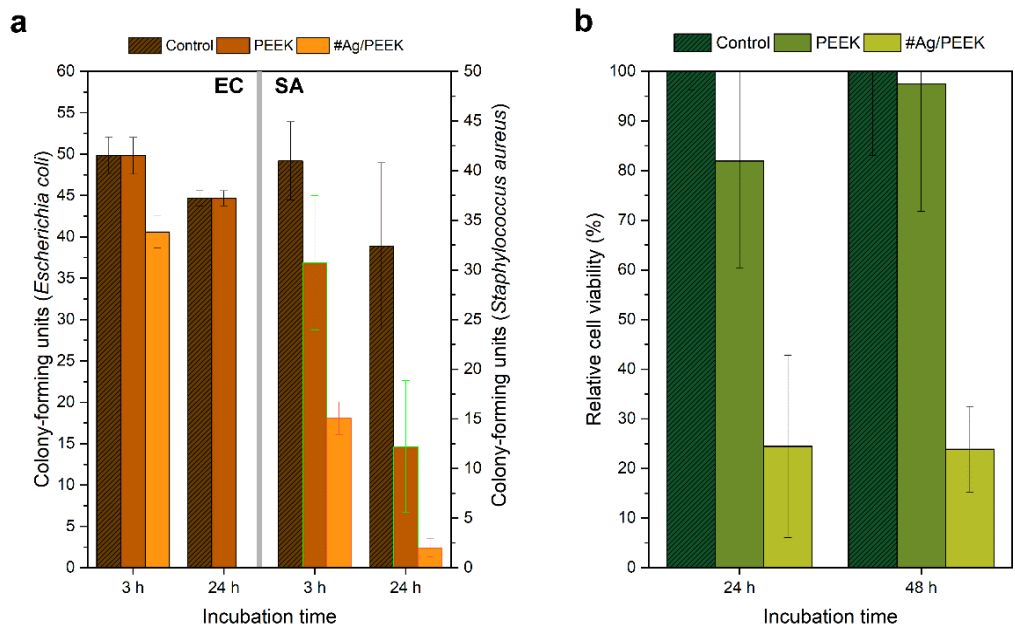

**Figure 6.** Results of (a) antibacterial and (b) cytotoxicity tests processed as graphs with standard deviation. #Ag/PEEK showed a strong antibacterial effect against *Escherichia coli* (left Y-axis) and *Staphylococcus aureus* (right Y-axis) and a significant cytotoxic effect on fibroblasts of the lungs.